\documentstyle{article}
\newcommand{\no}{\nonumber\\}
\newcommand{\be}{\begin{equation}}
\newcommand{\ee}{\end{equation}}
\newcommand{\ba}{\begin{eqnarray}}
\newcommand{\ea}{\end{eqnarray}}

\newcommand{\bi}[1]{\bibitem{#1}}

\date{}
\begin{document}

\centerline{{\bf DYNAMICAL LORENTZ SYMMETRY BREAKING FROM {\it 3+1}}}
\centerline{{\bf RENORMALIZABLE MODEL WITH WESS-ZUMINO INTERACTION}\footnote{Talk
given at XI International Workshop on High Energy Physics and Quantum Field Theory,
12-18 September 1996, Sankt-Petersburg}}

\vspace{4mm}

\centerline{{\bf A. A. Andrianov}
\footnote{E-mail:  andrianov1@phim.niif.spb.su}
\footnote{Supported by  RFBR (Grant No. 96-01-00535) and
by  INTAS (Grant No. 93-283-ext).}}
\centerline{\sl Department of Theoretical
Physics, University of Sankt Petersburg,}
\centerline{\sl 198904 Sankt Petersburg, Russia}

\medskip

\centerline{and}

\medskip

\centerline{{\bf R. Soldati}
\footnote{E-mail: soldati@bo.infn.it}
\footnote{Supported by INFN grant and by Italian MURST - quota 40/\%.}}
\centerline{\sl Dipartimento di Fisica "A. Righi", Universit\'a
di Bologna and Istituto Nazionale}
\centerline{\sl di Fisica Nucleare, Sezione di Bologna,
40126 Bologna, Italy}

\vspace{4mm}

 \noindent
{\bf Abstract}: \quad{\it
We study the renormalizable abelian vector-field models in the presence
of the Wess-Zumino
interaction with the pseudoscalar matter. The renormalizability is achieved
by supplementing the standard kinetic term of vector fields with higher 
derivatives.
The appearance  of fourth power of momentum in the vector-field propagator
leads to the super-renormalizable theory in which the $\beta$-function,
the vector-field renormalization constant and the anomalous mass dimension
are calculated exactly. It is shown that this model has the infrared
stable fixed point and its low-energy limit is non-trivial. 
The modified effective potential for the pseudoscalar matter 
leads to the occurrence of the quantum 
dynamical breaking of Lorentz symmetry.}

\vspace{4mm}

Anomalous gauge models might be consistently quantized: a first example
was provided by the  {\it 1+1} dimensional chiral Schwinger model [1].
The situation in the case of {\it 3+1} dimensional chiral gauge
theories is a very interesting but still open issue. The basic idea
to deal with [2-4], is to restore gauge invariance by means of some additional
quantized scalar fields. Here we aim to discuss 
the {\it 3+1} chiral massless abelian model as described by the lagrangian
$$
{\cal L}_0[A_\mu,\psi,\bar\psi]=-{1\over 4}F_{\mu\nu}F^{\mu\nu}
+\bar\psi\gamma^\mu\left\{i\partial_\mu + e A_\mu P_L\right\}\psi
+ \ \mbox{\rm gauge\ fixing}\ ,
$$
where $P_L\equiv (1/2)(\mbox{\bf 1}-\gamma_5)$ and which leads to the chiral
anomaly upon quantization, thereby breaking the classical invariance
under local gauge transformations of the left chiral sector.

To restore it, one might attempt to consider the gauge-group extension
\ba
 {\cal L}_0[A_\mu,\psi,\bar\psi]\longmapsto
{\cal L}[A_\mu,\psi,\bar\psi,\theta]&=&\-{1\over 4}F_{\mu\nu}F^{\mu\nu}
+\bar\psi\gamma^\mu\left\{i\partial_\mu + e (A_\mu + \partial_\mu
\theta)P_L\right\}\psi\no
&& + \  \mbox{\rm gauge\ fixing} + {\cal L}_{\mbox{\rm kin}}[\theta, A_\mu]\ ,
\ea
in which the so called (pseudo)scalar  
Wess-Zumino field indeed appears. Now the low-energy content of the above 
model is actually described by the effective lagrangian [5]
$$
{\cal L}_{\mbox{\rm eff}}={\cal L}_0[A_\mu,\psi,\bar\psi]+{e^3\over 
48\pi^2}\theta\tilde F_{\mu\nu}
F^{\mu\nu}+\ldots\ ,
$$
where the so called Wess-Zumino interaction arises.
It has been shown [6] that models of this kind described by the lagrangian
of eq.~(1) are indeed, by construction, BRST invariant: however there is a
serious conflict between power counting renormalizability vs. perturbative
unitarity.

Here, we would like to discuss the properties of the Wess-Zumino interaction,
which turn out to be quite relevant with respect to the above issues. 
The renormalizable abelian vector-field model 
(in the Euclidean space) we consider is
given by the lagrangian which contains the Wess-Zumino interaction
and the higher derivative kinetic term:
\ba
{\cal L}_{WZ}\ & =&\ {1\over 4M^2} \partial_{\rho}F_{\mu\nu}
\partial_{\rho}F_{\mu\nu}\  +\
{1\over 4} F_{\mu\nu} F_{\mu\nu}\ 
\ + \ {1\over 2\xi} (\partial_{\mu}A_{\mu})^2\no
&& + {1\over 2}\partial_{\mu}\theta \partial_{\mu}\theta\ +\
i{\kappa\over 2M} \theta
F_{\mu\nu} \tilde F_{\mu\nu}\ ,  
\ea
where $\tilde F_{\mu\nu}
 \equiv (1/2)\epsilon_{\mu\nu\rho\sigma} F_{\rho\sigma}$,
some suitable  dimensional scale $M$ is introduced,
 $\kappa$ and $\xi$ being the dimensionless coupling and gauge fixing 
parameter respectively. 

The Wess-Zumino interaction  can
be equivalently represented in the following form,
\be
\int d^4x\ {\kappa\over 2M}\ \theta\ F_{\mu\nu} \tilde F_{\mu\nu} \ =\
- \ \int d^4x\,{\kappa\over M}\, \partial_{\mu}\theta\, A_{\nu}
\tilde F_{\mu\nu}\ ,
\ee
when it is treated in the action. Therefore the pseudoscalar field
is involved into the dynamics only through its gradient
$\partial_{\mu}\theta(x)$ due to topological triviality of abelian
vector fields.

From the above lagrangian it is easy to derive the Feynman rules: namely,
the free vector field propagator reads
\be
D_{\mu\nu}(p)=-M^2{d_{\mu\nu}(p)\over p^2(p^2+M^2)}+{\xi\over p^2}
{p_\mu p_\nu\over p^2}\ ,
\ee
where $d_{\mu\nu}(p)\equiv -\delta_{\mu\nu}+(p_\mu p_\nu/p^2)$ is the 
transversal projector; the free scalar propagator is the usual 
$D(p)=(p^2)^{-1}$ and the vector-vector-scalar WZ-vertex 
turns out to be given by
\be
V_{\mu\nu}(p,q,r) = i(\kappa/M)\epsilon_{\mu\nu\rho\sigma}p_\rho q_\sigma\ ,
\qquad (p+q+r=0)
\ee
all momenta being incoming, $r$ being referred to the scalar field.
It is worthwhile to recall that the Fok space of asymptotic states, in the
Minkowskian formulation of
the present model, exhibits an indefinite metric structure. As a matter of fact,
from the algebraic identity
$$
{M^2\over p^2(p^2+M^2)}\ \equiv\ {1\over p^2}\ -\ {1\over p^2+M^2}\ ,
$$
it appears that negative norm states indeed are generated by the 
asymptotic  transversal
component  of vector field  with ghost-mass $M$; in addition, the 
longitudinal components of vector field 
give rise as well to negative norm states.
 
Let us now develop the power counting analysis of the superficial degree of 
divergence within the model. The number of loops is, as usual, $L= I_v+I_s-V+1$,
$I_{v(s)}$ being the number of vector (scalar) internal lines and $V$ the number
of vertices. Next we have $2V=2I_v+E_v$ and $V=2I_s+E_s$, where $E_{v(s)}$ is
the number of vector (scalar) external lines. As a consequence the overall
UV behaviour of a graph $G$ is provided by the exponent
\be
\omega (G)=\, 4L-4I_v-2I_s+2V-E_s-E_v\,=\,4-2E_v-E_s-2I_v+2I_s\ ,
\ee
and therefrom we see that the {\sl only} divergent graph corresponds to
\footnote{Actually the tadpole $E_s=I_v=1,\,I_s=0$ 
indeed vanishes owing to the tensorial 
structure of the WZ-vertex}  $I_s=1,\ I_v=1,\ E_s=0,\ E_v=2$ 
and it turns out to be the one
loop vector self-energy. Thus we conclude that the model is 
super-renormalizable. We notice that the number of external vector
lines has to be even. The computation of the divergent self-energy can
be done using dimensional regularization (in $2\omega$ dimensional Euclidean 
space) and gives
\be
\Pi^{(1)}_{\mu\nu}(p)\,=\,{1\over 16\epsilon}{\alpha\over \pi}p^2d_{\mu\nu}(p)
+ \hat\Pi^{(1)}_{\mu\nu}(p)\ ,
\ee
with $\epsilon\equiv 2-\omega,\,\alpha\equiv (\kappa^2/4\pi)$, 
while the finite part reads
\ba
\hat\Pi^{(1)}_{\lambda\nu}(p)\, & =&- {\alpha\over 16\pi}p^2d_{\lambda\nu}(p)
\times\left\{\ln{M^2\over 4\pi\mu^2} - \psi(2)+{2\over 3}
+{2\over 3}{p^2+M^2\over p^2}
\ln\left(1+{p^2\over M^2}\right)\right.\no
&&\left. - {M^2\over 3p^2}\left[1-{p^2+M^2\over p^2}
\ln\left(1+{p^2\over M^2}\right)\right]\right.\no
&& \left.-{p^2\over 3M^2}\left[1- {p^2+M^2\over
p^2}\ln\left(1+{p^2\over M^2}\right)+\ln {p^2\over M^2}\right]\right\}\ .
\ea
where $\mu$ denotes as usual the mass parameter in the 
dimensional regularization.
It follows therefore that the single countergraph to be added, in order to make 
finite the whole set of proper vertices, is provided by the 2-point 1PI
structure
\be
\Gamma_{\lambda\nu}^{(\mbox{\rm c.t.})}(p)\equiv -\left.\Pi^{(1)}_{\lambda\nu}(p)
\right|_{\mbox{\rm div}}\,=\,-{1\over 16}{\alpha\over \pi}p^2d_{\lambda\nu}(p)
\left[{1\over \epsilon}\,+\,F_1\left(\epsilon,{M^2\over 4\pi\mu^2}\right)
\right]\ ,
\ee
in which $F_1$ denotes the scheme-dependent finite part (when $\epsilon\to 0$)
of the countergraph.

As a result it is clear that we can write the renormalized lagrangian 
in the form
\ba
{\cal L}_{WZ}^{(\mbox{\rm ren})}\ & =&\ 
{1\over 4M_0^2} \partial_{\rho}F^{(0)}_{\mu\nu}
\partial_{\rho}F^{(0)}_{\mu\nu}\  +\
{1\over 4} F^{(0)}_{\mu\nu} F^{(0)}_{\mu\nu}\ 
\ + \ {1\over 2\xi_0} (\partial_{\mu}A^{(0)}_{\mu})^2\no
&& + {1\over 2}\partial_{\mu}\theta \partial_{\mu}\theta\ +\
{i\kappa_0\over 2M_0} \theta
F^{(0)}_{\mu\nu} \tilde F^{(0)}_{\mu\nu}\no
& =& {1\over 4M^2} \partial_{\rho}F_{\mu\nu}
\partial_{\rho}F_{\mu\nu}\  +\
{Z\over 4} F_{\mu\nu} F_{\mu\nu}\ 
\ + \ {1\over 2\xi} (\partial_{\mu}A_{\mu})^2\no
&& + {1\over 2}\partial_{\mu}\theta \partial_{\mu}\theta\ +\
i\mu^\epsilon{\kappa\over 2M} \theta
F_{\mu\nu} \tilde F_{\mu\nu}\ ,  
\ea
where the exact, due to super-renormalizability,
wave function renormalization constant $Z$ is provided by
\be
Z\ =\ c_0\left(\alpha,{M\over \mu};\epsilon\right)\ +\ 
{1\over \epsilon}c_1(\alpha)\ ;
\ee
here we can write, up to the one loop approximation, 
\ba
&& c_0\left(\alpha,{M\over \mu};\epsilon\right)\ =\ 1\ -\ {\alpha\over
16\pi}F_1\left(\epsilon,{M^2\over 4\pi\mu^2}\right)\ +\ {\cal O}(\alpha^2)\ ,\no
&& c_1(\alpha)\ =\ {-\alpha\over 16\pi}\ .
\ea
Moreover the relationships between bare and renormalized quantities
turn out to be the following,
\be
A_\mu^{(0)}\ =\ \sqrt{Z} A_\mu\ ,\quad
 M_0\ =\ \sqrt{Z} M\ , \quad \xi_0\ =\ Z \xi\ ,\quad
 \kappa_0\ =\ \mu^\epsilon {\kappa\over \sqrt{Z}}\ ,
\quad \alpha_0\ =\ {\alpha\over Z}\ .
\ee
In particular, from the Laurent expansion of eq.~(13), we can write
\be
\kappa_0\ =\ \mu^\epsilon\left\{a_0\left(\kappa,{M\over \mu};\epsilon\right)\ 
+\ {1\over \epsilon} a_1(\kappa)\right\}\ ,
\ee
with
\ba
&& a_0\left(\kappa,{M\over \mu};\epsilon\right)\ =\ 
\kappa+{\kappa^3\over 128\pi^2}F_1\left(\epsilon,{M^2\over 4\pi\mu^2}\right)
+{\cal O}(\kappa^5)\ ,\no
&& a_1(\kappa)\ =\ {\kappa^3\over 128\pi^2}\ .
\ea

This entails that, within this model, we can solve the renormalization group
equations (RGE) in the minimal subtraction (MS) scheme $F_1\equiv 0$: namely,
\be
\mu{\partial\kappa\over \partial\mu}\ =\ -\epsilon\kappa-a_1(\kappa)+
\kappa{d\over d\kappa}a_1(\kappa)\ ,
\ee
to get the exact MS prescription for $\beta$-function
\be
\beta (\kappa)\ =\ {\kappa^3\over 64\pi^2}\ ,\quad
\beta (\alpha)\ =\ {\alpha^2\over 8\pi}\ ,
\ee
which tells us, as expected, that $\alpha=0$ is an IR stable fixed point.
Therefrom it follows that we can integrate eq.~(16) and determine the 
exact behaviour of running coupling  in perturbation theory
\be
\alpha(\mu)\ =\ {\alpha(\mu_0)\over 1-[\alpha(\mu_0)/8\pi]\ln (\mu/\mu_0)}\ .
\ee
Furthermore, from eqs~(13) and within the MS prescription, 
it is straightforward to determine the remaining RG coefficients to be
\ba
&& \gamma_M\ \equiv\ {1\over 2}\mu{\partial\ln M^2\over \partial\mu}\ =\
{-\alpha\over 16\pi}\ , \no
&& \gamma_d\ \equiv\ {1\over 2}\mu{\partial\ln Z\over \partial\mu}\ =\
{\alpha\over 8\pi}\ , \no
&& \gamma_\xi\ \equiv\ \mu{\partial\ln\xi\over \partial\mu}\ =\
{-\alpha\over 4\pi}\ . 
\ea
In conclusion, we are able to summarize the asymptotic behaviour
of the ghost-mass parameter $M$ and of the gauge-fixing parameter $\xi$ at
large distances, where perturbation theory is reliable in the model
we are considering and within the MS renormalization scheme. Actually,
if we set $s\equiv (\mu/\mu_0)$, we can easily derive
\ba
&& \bar\alpha(s;\alpha)\ =\ {\alpha\over 1-(\alpha/8\pi)\ln s}\
\buildrel s\to 0 \over \sim\ -{8\pi\over \ln s}\ , \no
&& \bar M(s;M,\alpha)\ =\ M\sqrt{1-{\alpha\over 8\pi}\ln s}\
\buildrel s\to 0 \over \sim\ M\sqrt{{\alpha|\ln s|\over 8\pi}}\ ,\no
&& \bar\xi (s;\xi,\alpha)\ =\ \xi+\ln\left(1-{\alpha\ln s\over 8\pi}\right)\
\buildrel s\to 0 \over \sim\ \xi+2\ln\left({\alpha\over 4\pi}|\ln s|\right)\ 
\ea
showing that longitudinal as well as ghost-like transversal degrees of freedom of vector
fields decouple at small momenta where perturbation 
theory has to be trusted. Owing to this asymptotic decoupling of negative 
norm states, within the domain of validity of perturbation theory, the present 
super-renormalizable model might be referred to as {\sl asymptotically unitary}.

We are ready now to discuss a further very interesting feature of this
simple but non trivial model: the occurrence of the radiative
Coleman-Weinberg [7]  breaking,
at the quantum level,
of the $SO(4)$-symmetry in the Euclidean version, or the 
$O(3,1)^{++}$ space-time symmetry in the Minkowskian case.
As a matter of fact, we shall see in the following that the effective
potential for the pseudoscalar field $\theta$ exhibits non
trivial true minima and, consequently, some privileged direction has
to be fixed by boundary conditions, in order to specify the vacuum of the model.
More interesting, those non trivial minima lie within the perturbative domain.
Since we are looking for the effective potential of the pseudoscalar field,
we are allowed to ignore the renormalization constant $Z(\epsilon)$ and restart
from the classical action in four dimensions: namely,
\ba
{\cal A}_{WZ}[A_\mu,\theta] 
&=& \int d^4x\left\{{\rho\over 4M_*^2} \partial_{\lambda}F_{\mu\nu}(x)
\partial_{\lambda}F_{\mu\nu}(x)+
{1\over 4} F_{\mu\nu}(x) F_{\mu\nu}(x) 
+ {1\over 2\xi} [\partial_{\mu}A_{\mu}(x)]^2\right.\no
&& \left.+ {1\over 2}\partial_{\mu}\theta (x) \partial_{\mu}\theta (x) +
{i\over 2M_*} \theta (x)
F_{\mu\nu}(x) \tilde F_{\mu\nu}(x)\right\}\ ,  
\ea
in which we introduce the suitable parametrization $M_*\equiv (M/\alpha),\ 
\rho\equiv (M_*^2/M^2)$. The  generating functional 
for pseudoscalar
background field is defined as
\be
{\cal Z}[\theta]\ \equiv\ {\cal N}^{-1}\int [{\cal D}A_\mu]\
\exp\left\{-{\cal A}_{WZ}[A_\mu,\theta]\right\}\ .
\ee
The classical field configurations $\bar A_\mu (x)$ are the solutions of the
Euler-Lagrange equations
\be
{\delta {\cal A}_{WZ}[A_\mu,\theta]\over \delta A_\mu (x)}\ =\
K_{\mu\nu}[\theta]\bar A_\nu (x)\ =\ 0\ ,
\ee
with ($\triangle\equiv \partial_\mu\partial_\mu$)
\be
K_{\mu\nu}[\theta]\ \equiv\ 
\left(\rho{\triangle\over M_*^2}-1\right)
\left(\delta_{\mu\nu}\triangle-\partial_\mu\partial_\nu\right)
-\ {1\over \xi}\partial_\mu\partial_\nu + {1\over M_*}
\epsilon_{\lambda\mu\sigma\nu}\partial_\lambda\theta (x)
(-i\partial_\sigma)\ ,
\ee
being an elliptic invertible local differential operator. Therefore, if we set
$a_\mu (x)\equiv A_\mu (x)-\bar A_\mu (x)$, we eventually obtain
\be
{\cal Z}[\theta]\ \equiv\ {\cal N}^{-1}
\exp\left\{-{\cal A}_{WZ}[\bar A_\mu,\theta]\right\}\times
\left(\mbox{\rm det}\parallel K_{\mu\nu}[\theta]\parallel\right)^{-1/2}\ ,
\ee
with ${\cal N}={\cal Z}[\theta=0]$.

In order to evaluate the effective 
potential it is more convenient to consider the dimensionless operator
\be
{\cal K}_{\mu\nu}[\theta]\ \equiv\ (1/M_*^2)K_{\mu\nu}[\theta]
=\ -\top_{\mu\nu}{\triangle\over M_*^2}
\left(\rho{\triangle\over M_*^2}-1\right)
-{1\over \xi}{\triangle\over M_*^2}\ell_{\mu\nu}
\ +{1\over M_*}\epsilon_{\mu\nu\lambda\sigma}\eta_\lambda (x)
(-i\partial_\sigma)\ ,
\ee
where we have set
\be
 \top_{\mu\nu}\ \equiv\ -\delta_{\mu\nu}\ +\ {\partial_\mu\partial_\nu\over 
\triangle}\ ,\quad
\ell_{\mu\nu}\ \equiv\ {\partial_\mu\partial_\nu\over \triangle}\ ,\quad
 \eta_\mu (x)\ \equiv\ (1/M_*^2)\partial_\mu\theta (x)\ .
\ee
We want to evaluate the determinant of eq.~(25) for {\sl constant} 
dimensionless vector $\eta_\mu$; to this aim we can rewrite the relevant
operator into the form
\be
{\cal K}_{\mu\nu}(\eta)\ \equiv\ 
{\triangle\over M_*^2}
\left(\tau_{\mu\nu}- 
{1\over \xi}\ell_{\mu\nu}\right)
+ {\cal E}_{\mu\nu}(\eta)\ ,
\ee
with
\be
{\cal E}_{\mu\nu}(\eta)\ \equiv\ 
{1\over M_*}\epsilon_{\mu\nu\lambda\sigma}\eta_\lambda
(-i\partial_\sigma)\ ,\quad
\tau_{\mu\nu}\ \equiv\ \top_{\mu\nu}
\left(1-\rho{\triangle\over M_*^2}\right)\ .
\ee
From the conjugation property
\be
\left({\cal E}^\dagger\right)_{\mu\nu}\ =\ -{\cal E}_{\mu\nu}\ ,
\ee
it follows that
\be
\left({\cal K}^\dagger[\eta]\right)_{\mu\nu}\ =\ 
\left({\cal K}[-\eta]\right)_{\mu\nu}\ ,
\ee
which shows that the the relevant operator is {\sl normal}.
As a consequence, after compactification of the
Euclidean space, we can safely define its complex power [8] and its
determinant [9] by means of the $\zeta$-function technique:
namely,
\be
\mbox{\rm det}\left\Vert{\cal K}[\eta]\right\Vert\ =\ 
\left( \mbox{\rm det}\left\Vert{\cal K}[\eta]{\cal K}^\dagger 
[\eta]\right\Vert\right)^{1/2}
\equiv\ \left.\exp\left\{-{1\over 2}{d\over ds}\zeta_H (s;\eta)
\right\}\right\vert_{s=0}\ ,
\ee
where we have set\footnote{the same regularized determinant is
obtained by considering $H^\prime[\eta]\equiv {\cal K}^\dagger[\eta]{\cal K} 
[\eta]$.}
\ba
\left(H[\eta]\right)_{\mu\nu}\ &\equiv&\ 
\left({\cal K}[\eta]\right)_{\mu\lambda}
\left({\cal K}^\dagger [\eta]\right)_{\lambda\nu}\ , \\
\zeta_H (s;\eta)\ &\equiv&\ \mbox{\rm Tr}\left(H[\eta]\right)^{-s}\ .
\ea
 
After some straightforward calculations, we can definitely obtain  
\ba
{\cal W}[\eta_\mu ,\rho]=-\ln{\cal Z}[\eta_\mu,\rho]&\equiv&
{\cal A}_{WZ}[\bar A_,\eta,\rho]-
{\cal A}_{WZ}[\bar A_,\eta=\rho=0]\no
&&-{1\over 4}{d\over ds}\zeta_H (s=0;\eta,\rho)+
{1\over 4}{d\over ds}\zeta_{h_0} (s=0)\ ,
\ea
in which
\be
\zeta_H (s;\eta,\rho)=2(\mbox{\rm vol})_4\int {d^4 p\over (2\pi)^4}
\left\{(p^2)^2\left(1+\rho{p^2\over M_*^2}\right)^2+M_*^2\left(
(\eta\cdot p)^2-\eta^2 p^2\right)\right\}^{-s}\ ,
\ee
while, obviously, $\zeta_{h_0} (s)=\zeta_H (s;\eta=\rho=0)$.
The effective potential for constant $\eta_\mu$ appears
eventually to be expressed as
\be
{\cal V}_{\mbox{\rm eff}}(\eta,\rho)\ \equiv\  (\mbox{\rm vol})_4^{-1}\left\{
-{1\over 4}{d\over ds}\zeta_H (s=0;\eta,\rho)+
{1\over 4}{d\over ds}\zeta_{H_0} (s=0)\right\}\ ,
\ee
and therefore we have to compute carefully the integral in eq.~(36).
To this aim, it is convenient to select a coordinate system in which
\be
p_\mu\ =\ (\mbox{\bf p},p_4)\ ,\quad p_4\ =\ {\eta\cdot p\over 
\sqrt{\eta^2}}\ ,
\ee
so that, after rescaling variables to
$\mbox{\bf v}=(\mbox{\bf p}/M_*),\ y=(p_4/M_*)$, we obtain
\ba
&&(\mbox{\rm vol})_4^{-1}\zeta_H (s;\eta,\rho)\ 
=\ {4\over (2\pi)^4\Gamma (s)}\times\no
&&\int_0^\infty d\tau\ \tau^{s-1}\int_0^\infty dy\int d^3x\
\exp\left\{-\tau\left(\mbox{\bf v}^2+y^2\right)^2\left(1
+\rho\left(\mbox{\bf v}^2+
y^2\right)\right)^2+\tau\eta^2\mbox{\bf v}^2\right\}\ .
\ea
A straightforward calculation leads eventually to the following integral
representation \footnote{We notice that, from the integral 
representation (40) for Re $s<1$, it turns out that $\zeta_{H_0}(s)$ is
regularized to zero.}  [10]: 
\be
(\mbox{\rm vol})_4^{-1}\zeta_H (s;\eta,\rho)={(\eta^2)^{2-2s}\over 8\pi^2}
\int_0^\infty dt\ {t^{1-2s}\over \left(1-\rho\eta^2 t\right)^{2s}}\
_2F_1\left({3\over 2},s;2;{-1\over t\left(1-\rho\eta^2 t\right)^2}\right)\ .
\ee

Let us first analyze the case $\rho=0$, which corresponds to the low-energy
unitary regime; in this limit, the integration in the previous formula can be
performed explicitely ($1<$ Re $s <(7/4)$) to yield
\be
(\mbox{\rm vol})_4^{-1}\zeta_H (s;\eta,\rho=0)=
{(\eta^2)^{2-2s}\over 16\pi^2\sqrt\pi}{2^{4s-4}\over (s-1)}
{\Gamma[s-(1/2)]\Gamma[(7/2)-2s]\over \Gamma[(5/2)-s]}\ .
\ee
In the present case $\rho\to 0$, the effective potential for constant 
$\eta_\mu$ within the $\zeta$-function regularization is given by
\be
 {\cal V}_{\mbox{\rm eff}}(\eta,\rho=0)=
-(\mbox{\rm vol})_4^{-1}
{1\over 4}{d\over ds}\zeta_H (s=0;\eta,\rho=0)=
{5z^2\over 64\pi^2}\left(2\ln z +{7\over 15}\right)\ ,
\ee
where $z\equiv (\eta^2/4)$. We see that the stable $O(4)$-degenerate
non trivial minima correspond to the symmetry breaking value
\be
\ln z_{SB}\ +\ {11\over 15}\ =\ 0\ ,\quad z_{SB}\ =\ \exp\{-0.7333\ldots\}\ .
\ee
We remark that the above result, within the $\zeta$-function regularization,
actually reproduces our previous calculation [5] using
large momenta cutoff regularization. To be more precise, eq.~(42) indeed
corresponds to a specific choice of the subtraction terms in the large momenta
cutoff method, something that we could call {\sl minimal subtraction for the
effective potential}.
As a matter of fact we recall
that, in general, the $\zeta$-regularized functional determinants
of elliptic invertible normal operators are defined
up to local polynomials of the background fields.

\bigskip

To sum up, we can draw the following conclusions: \\
$i)$\ in the {\it 3+1}
dimensional abelian vector-scalar model with the Wess-Zumino interaction,
the renormalization group behaviour allows to reconcile, in
some sense, perturbative renormalizability and unitarity, 
in the asymptotic low momenta domain where perturbation theory is reliable.\\
$ii)$ \ It is obviously very interesting to investigate
whether a similar feature still holds, within the fully realistic models 
involving chiral fermions.\\
$ iii)$ \ Consistent gauge invariant quantization, if any, leads unavoidably
to the quantum dynamical breaking of the Lorentz symmetry; 
this phenomenon has been 
also noticed [11] in the framework of  {\it 2+1} dimensional
Chern-Simons theories. The origin of this symmetry breaking is absolutely similar to the
Coleman-Weinberg mechanism and is related to the renormalization of one-loop
divergences [12].
What a  physical meaning could be  eventually hidden behind this phenomenon 
will be clarified elsewhere.

\end{document}